\documentclass[fleqn]{llncs}
\usepackage{tepis}

\title{Thread Extraction for\\ Polyadic Instruction Sequences%
       \thanks{This research has been partly carried out in the
               framework of the Jacquard-project Symbiosis, which is
               funded by the Netherlands Organisation for Scientific
               Research (NWO).}}
\author{J.A. Bergstra \and C.A. Middelburg}
\institute{Informatics Institute, Faculty of Science,
           University of Amsterdam, \\
           Science Park~107, 1098~XG~Amsterdam, the Netherlands \\
           \email{J.A.Bergstra@uva.nl,C.A.Middelburg@uva.nl}}

\begin{document}

\maketitle

\begin{abstract}
In this paper, we study the phenomenon that instruction sequences are
split into fragments which somehow produce a joint behaviour.
In order to bring this phenomenon better into the picture, we formalize
a simple mechanism by which several instruction sequence fragments can
produce a joint behaviour.
We also show that, even in the case of this simple mechanism, it is a
non-trivial matter to explain by means of a translation into a single
instruction sequence what takes place on execution of a collection of
instruction sequence fragments.
\begin{classcode}
68N15; 68Q05; 68Q55
\end{classcode}
\end{abstract}

\section{Introduction}
\label{sect-intro}

With the work presented in this paper, we carry on a line of research
with which a start was made in~\cite{BL02a}.
This line of research is concerned with sequential programs that take
the form of instruction sequences.
Its working hypothesis is that instruction sequence is a central notion
of computer science, which merits investigation for its own sake.

An instruction sequence is considered to produce on execution a
behaviour to be controlled by some execution environment.
This behaviour proceeds by performing steps in a sequential fashion.
Each step performed actuates the processing of an instruction by the
execution environment in question.
A reply returned by this execution environment at completion of the
processing of the instruction determines how the behaviour proceeds
further.

The following phenomenon presents itself: instruction sequences are
split into fragments which somehow produce a joint behaviour.
The objective of this paper is to bring this phenomenon better into the
picture.
To achieve this, we formalize a simple mechanism by which
several instruction sequence fragments can produce a joint behaviour.
We show that, even in the case of this simple mechanism, it is a
non-trivial matter to explain by means of a translation into a single
instruction sequence what takes place on execution of a collection of
instruction sequence fragments.

The question is how a joint behaviour of the fragments in a collection
of fragments is achieved.
The view of this matter is that there can only be a single fragment
being executed at any stage, but the fragment in question may make any
fragment in the collection the one being executed by means of a special
instruction for switching over execution to another fragment.
This does not fit in very well with the conception that the collection
of fragments constitutes a sequential program.
To our knowledge, a theoretical understanding of this matter has not yet
been developed.
This has motivated us to take up this topic.

The principal reason for splitting instruction sequences into fragments
is that the execution environment at hand sets bounds to the size of
instruction sequences.
In the past, the phenomenon occurred explicitly in many software systems.
At present, it often occurs rather implicitly, e.g.\ on execution of
programs written in contemporary object-oriented programming languages,
such as Java~\cite{AG96a} and C\#~\cite{BH04a}, classes are loaded as
they are needed.
The mechanisms in question are improvements upon the simple mechanism
considered in this paper, but they are also much more complicated.
We believe that it is useful to consider the simple mechanism prior to
the more complicated ones.

The instruction sequences taken for fragments are called polyadic
instruction sequences in this paper.
We introduce polyadic instruction sequences in the setting of program
algebra~\cite{BL02a}.
The starting-point of program algebra is the perception of a program as
a single-pass instruction sequence, i.e.\ a finite or infinite sequence
of instructions of which each instruction is executed at most once and
can be dropped after it has been executed or jumped over.
This perception is simple, appealing, and links up with practice.

The behaviours produced by instruction sequences on execution are
modelled by threads as considered in basic thread algebra~\cite{BL02a}.%
\footnote
{In~\cite{BL02a}, basic thread algebra is introduced under the name
 basic polarized process algebra.}
We take the view that the possible joint behaviours produced by polyadic
instruction sequences on execution are threads as considered in basic
thread algebra as well.
In a system that provides an execution environment for polyadic
instruction sequences, a polyadic instruction sequence must be loaded in
order to become the one being executed.
Hence, making a polyadic instruction sequence the one being executed can
be looked upon as loading it for execution.

In~\cite{BL02a}, a hierarchy of program notations rooted in program
algebra is presented.
Included in this hierarchy are very simple program notations which are
close to existing assembly languages up to and including simple program
notations that support structured programming by offering a rendering of
conditional and loop constructs.
All of these program notations are referred to in this paper, but only
one of them is actually used.
That program notation is introduced under the name \PGLD\
in~\cite{BL02a}.

This paper is organized as follows.
First, we review basic thread algebra and program algebra
(Sections~\ref{sect-BTA} and~\ref{sect-PGA}).
After that, we give an overall picture of the hierarchy of program
notations rooted in program algebra and present the program notation
\PGLD\ (Sections~\ref{sect-hierarchy} and~\ref{sect-PGLD}).
Next, we introduce polyadic instruction sequences in the setting of
program algebra, explain the possible joint behaviours of a collection
of polyadic instruction sequences using basic thread algebra, and give
an example of the use of polyadic instruction sequences
(Sections~\ref{sect-polyadic-inseqs} and~\ref{sect-example}).
Following this, we extend basic thread algebra to allow for threads to
make use of services and give a description of instruction register file
services (Sections~\ref{sect-TSU} and~\ref{sect-IRF}).
After that, we show that, for each possible joint behaviour of a
collection of polyadic instruction sequences, a single instruction
sequence can be synthesized from the collection of polyadic instruction
sequences that produces on execution essentially the behaviour in
question by making use of an instruction register file service
(Section~\ref{sect-pg-synthesis}).
Finally, we make some concluding remarks (Section~\ref{sect-concl}).

In this paper, we only give brief summaries of program algebra and basic
thread algebra.
Comprehensive introductions, including examples, can be
found in~\cite{BL02a,PZ06a}.

\section{Basic Thread Algebra}
\label{sect-BTA}

In this section, we review \BTA\ (Basic Thread Algebra).
\BTA\ is a form of process algebra which is concerned with the behaviour
that sequential programs produce on execution.
Those behaviours are called \emph{threads}.

In \BTA, it is assumed that fixed but arbitrary finite sets $\BAct$ and
$\IAct$ with $\BAct \inter \IAct = \emptyset$ and $\Tau \in \IAct$ have
been given.
The members of $\BAct$ are called \emph{basic actions} and the members
of $\IAct$ are called \emph{internal actions}.
The members of $\Act$ are referred to as \emph{actions}.
In previous work, we take in essence the singleton set $\set{\Tau}$ for
$\IAct$.
The generalization made here permits internal actions with differences
relevant for analysis to be distinguished.

The operational intuition is that a thread has an execution environment
which processes each action performed by the thread.
A thread performs actions in a sequential fashion.
Upon each action performed, a reply from the execution environment of
the thread determines how it proceeds.
The possible replies are $\True$ and $\False$.
Performing an internal action, always leads to the reply $\True$.

Although \BTA\ is one-sorted, we make this sort explicit.
The reason for this is that we will extend \BTA\ with an additional sort
in Section~\ref{sect-TSU}.

\BTA\ has one sort: the sort $\Thr$ of \emph{threads}.
To build terms of sort $\Thr$, \BTA\ has the following constants and
operators:
\begin{itemize}
\item
the \emph{inaction} constant $\const{\DeadEnd}{\Thr}$;
\item
the \emph{termination} constant $\const{\Stop}{\Thr}$;
\item
for each $a \in \Act$, the binary \emph{postconditional composition}
operator $\funct{\pcc{\ph}{a}{\ph}}{\linebreak[2]\Thr \cp \Thr}{\Thr}$.
\end{itemize}
We assume that there are infinitely many variables of sort $\Thr$,
including $x,y,z$.
Terms of sort $\Thr$ are built as usual (see e.g.~\cite{ST99a,Wir90a}).
We use infix notation for the postconditional composition operator.
We introduce \emph{basic action prefixing} as an abbreviation:
$a \bapf p$ abbreviates $\pcc{p}{a}{p}$.

The thread denoted by a closed term of the form $\pcc{p}{a}{q}$ will
first perform $a$, and then proceed as the thread denoted by $p$
if the reply from the execution environment is $\True$ and proceed as
the thread denoted by $q$ if the reply from the execution environment is
$\False$.
The threads denoted by $\DeadEnd$ and $\Stop$ will become inactive and
terminate successfully, respectively.
A thread is inactive if it is neither capable of performing any action
nor capable of terminating successfully.

\BTA\ has only one axiom.
This axiom is given in Table~\ref{axioms-BTA}.%
\begin{table}[!t]
\caption{Axiom of \BTA}
\label{axioms-BTA}
\begin{eqntbl}
\begin{axcol}
\pcc{x}{\iota}{y} = \pcc{x}{\iota}{x}                    & \axiom{T1}
\end{axcol}
\end{eqntbl}
\end{table}
In this table, $\iota$ stands for an arbitrary member of $\IAct$.

Notice that each closed \BTA\ term denotes a thread that will become
inactive or terminate after it has performed finitely many actions.
Infinite threads can be described by guarded recursion.

A \emph{guarded recursive specification} over \BTA\ is a set of
recursion equations $E = \set{X = t_X \where X \in V}$, where $V$ is a
set of variables of sort $\Thr$ and each $t_X$ is a \BTA\ term of the
form $\DeadEnd$, $\Stop$ or $\pcc{t}{a}{t'}$ with $t$ and $t'$ that
contain only variables from $V$.
We write $\vars(E)$ for the set of all variables that occur in $E$.
We are only interested in models of \BTA\ in which guarded recursive
specifications have unique solutions, such as the projective limit model
of \BTA\ presented in~\cite{BB03a}.
A thread that is the solution of a finite guarded recursive
specification over \BTA\ is called a \emph{finite-state} thread.

For each guarded recursive specification $E$ and each $X \in \vars(E)$,
we introduce a constant $\rec{X}{E}$ of sort $\Thr$ standing for the
unique solution of $E$ for $X$.
The axioms for these constants are given in Table~\ref{axioms-REC}.%
\begin{table}[!t]
\caption{Axioms for guarded recursion}
\label{axioms-REC}
\begin{eqntbl}
\begin{saxcol}
\rec{X}{E} = \rec{t_X}{E} & \mif X \!=\! t_X \in E       & \axiom{RDP}
\\
E \Implies X = \rec{X}{E} & \mif X \in \vars(E)          & \axiom{RSP}
\end{saxcol}
\end{eqntbl}
\end{table}
In this table, we write $\rec{t_X}{E}$ for $t_X$ with, for all
$Y \in \vars(E)$, all occurrences of $Y$ in $t_X$ replaced by
$\rec{Y}{E}$.
$X$, $t_X$ and $E$ stand for an arbitrary variable of sort $\Thr$, an
arbitrary \BTA\ term of sort $\Thr$ and an arbitrary guarded recursive
specification over \BTA, respectively.
Side conditions are added to restrict what $X$, $t_X$ and $E$ stand for.

We will use the following abbreviation: $a^\omega$, where
$a \in \Act$, abbreviates $\rec{X}{\set{X = a \bapf X}}$.

We will write \BTA+\REC\ for \BTA\ extended with the constants for
solutions of guarded recursive specifications and axioms RDP and RSP.

Closed terms of sort $\Thr$ from the language of \BTA+\REC\ that denote
the same infinite thread cannot always be proved equal by means of the
axioms of \BTA+\REC.
We introduce \AIP\ (Approximation Induction Principle) to remedy this.
\AIP\ is based on the view that two threads are identical if their
approximations up to any finite depth are identical.
The approximation up to depth $n$ of a thread is obtained by cutting it
off after performing a sequence of actions of length $n$.
In \AIP, the approximation up to depth $n$ is phrased in terms of the
unary \emph{projection} operator $\funct{\projop{n}}{\Thr}{\Thr}$.
\AIP\ and the axioms for the projection operators are given in
Table~\ref{axioms-AIP}.%
\begin{table}[!t]
\caption{Approximation induction principle}
\label{axioms-AIP}
\begin{eqntbl}
\begin{axcol}
\AND{n \geq 0} \proj{n}{x} = \proj{n}{y} \Implies x = y & \axiom{AIP} \\
\proj{0}{x} = \DeadEnd                                  & \axiom{P0} \\
\proj{n+1}{\Stop} = \Stop                               & \axiom{P1} \\
\proj{n+1}{\DeadEnd} = \DeadEnd                         & \axiom{P2} \\
\proj{n+1}{\pcc{x}{a}{y}} =
                      \pcc{\proj{n}{x}}{a}{\proj{n}{y}} & \axiom{P3}
\end{axcol}
\end{eqntbl}
\end{table}
In this table, $a$ stands for an arbitrary member of $\Act$.

We will write \BTA+\REC+\AIP\ for \BTA+\REC\ extended with the
projection operators and the axioms from Table~\ref{axioms-AIP}.

\section{Program Algebra}
\label{sect-PGA}

In this section, we review \PGA\ (ProGram Algebra).
The perception of a program as a single-pass instruction sequence is the
starting-point of \PGA.

In \PGA, it is assumed that a fixed but arbitrary set $\BInstr$ of
\emph{basic instructions} has been given.
\PGA\ has the following \emph{primitive instructions}:
\begin{itemize}
\item
for each $a \in \BInstr$, a \emph{plain basic instruction} $a$;
\item
for each $a \in \BInstr$, a \emph{positive test instruction} $\ptst{a}$;
\item
for each $a \in \BInstr$, a \emph{negative test instruction} $\ntst{a}$;
\item
for each $l \in \Nat$, a \emph{forward jump instruction} $\fjmp{l}$;
\item
a \emph{termination instruction} $\halt$.
\end{itemize}
We write $\PInstr$ for the set of all primitive instructions.

The intuition is that the execution of a basic instruction $a$ produces
either $\True$ or $\False$ at its completion.
In the case of a positive test instruction $\ptst{a}$, $a$ is executed
and execution proceeds with the next primitive instruction if $\True$ is
produced.
Otherwise, the next primitive instruction is skipped and execution
proceeds with the primitive instruction following the skipped one.
If there is no next instruction to be executed, inaction occurs.
In the case of a negative test instruction $\ntst{a}$, the role of the
value produced is reversed.
In the case of a plain basic instruction $a$, execution always proceeds
as if $\True$ is produced.
The effect of a forward jump instruction $\fjmp{l}$ is that execution
proceeds with the $l$-th next instruction.
If $l$ equals $0$ or the $l$-th next instruction does not exist, then
$\fjmp{l}$ results in inaction.
The effect of the termination instruction $\halt$ is that execution
terminates.

\PGA\ has the following constants and operators:
\begin{itemize}
\item
for each $u \in \PInstr$, an \emph{instruction} constant $u$\,;
\item
the binary \emph{concatenation} operator $\ph \conc \ph$\,;
\item
the unary \emph{repetition} operator $\ph\rep$\,.
\end{itemize}
We assume that there are infinitely many variables, including $x,y,z$.
Terms are built as usual.
We use infix notation for the concatenation operator and postfix
notation for the repetition operator.

A closed \PGA\ term is considered to denote a non-empty, finite or
periodic infinite sequence of primitive instructions.%
\footnote
{An infinite sequence is periodic if the set of all its subsequences is
 finite.}
Closed \PGA\ terms are considered equal if they denote the same
instruction sequence.
The axioms for instruction sequence equivalence are given in
Table~\ref{axioms-PGA}.%
\begin{table}[!t]
\caption{Axioms of \PGA}
\label{axioms-PGA}
\begin{eqntbl}
\begin{axcol}
(x \conc y) \conc z = x \conc (y \conc z)              & \axiom{PGA1} \\
(x^n)\rep = x\rep                                      & \axiom{PGA2} \\
x\rep \conc y = x\rep                                  & \axiom{PGA3} \\
(x \conc y)\rep = x \conc (y \conc x)\rep              & \axiom{PGA4}
\end{axcol}
\end{eqntbl}
\end{table}
In this table, $n$ stands for an arbitrary natural number greater than
$0$.
For each \PGA\ term $P$,
the term $P^n$ is defined by induction on $n$ as follows: $P^1 = P$ and
$P^{n+1} = P \conc P^n$.
The equation $X\rep = X \conc X\rep$ is derivable.
Each closed \PGA\ term is derivably equal to one of the form $P$ or
$P \conc Q\rep$, where $P$ and $Q$ are closed \PGA\ terms in which the
repetition operator does not occur.

Each closed \PGA\ term $P$ is considered to denote an instruction
sequence of which the behaviour is a finite-state thread, called the
\emph{thread produced by} $P$.
The set $\BInstr$ of basic instructions is taken for the set $\BAct$ of
basic actions.
The \emph{thread extraction} operation $\extr{\ph}$ determines, for each
closed \PGA\ term $P$, a finite guarded recursive specification over
\BTA\ that defines the thread produced by $P$.
The thread extraction operation is defined by the equations given in
Table~\ref{axioms-thread-extr} (for $a \in \BInstr$, $l \in \Nat$ and
$u \in \PInstr$)%
\begin{table}[!t]
\caption{Defining equations for thread extraction operation of \PGA}
\label{axioms-thread-extr}
\begin{eqntbl}
\begin{eqncol}
\extr{a} = a \bapf \DeadEnd \\
\extr{a \conc x} = a \bapf \extr{x} \\
\extr{\ptst{a}} = a \bapf \DeadEnd \\
\extr{\ptst{a} \conc x} =
\pcc{\extr{x}}{a}{\extr{\fjmp{2} \conc x}} \\
\extr{\ntst{a}} = a \bapf \DeadEnd \\
\extr{\ntst{a} \conc x} =
\pcc{\extr{\fjmp{2} \conc x}}{a}{\extr{x}}
\end{eqncol}
\qquad
\begin{eqncol}
\extr{\fjmp{l}} = \DeadEnd \\
\extr{\fjmp{0} \conc x} = \DeadEnd \\
\extr{\fjmp{1} \conc x} = \extr{x} \\
\extr{\fjmp{l+2} \conc u} = \DeadEnd \\
\extr{\fjmp{l+2} \conc u \conc x} = \extr{\fjmp{l+1} \conc x} \\
\extr{\halt} = \Stop \\
\extr{\halt \conc x} = \Stop
\end{eqncol}
\end{eqntbl}
\end{table}
and the rule that $\extr{\fjmp{l} \conc x} = \DeadEnd$ if $\fjmp{l}$ is
the beginning of an infinite chain of forward jumps.
This rule is formalized in e.g.~\cite{BM07g}.

The behaviour of each closed \PGA\ term, is a thread that is definable
by a finite guarded recursive specification over \BTA.
The other way round, each finite guarded recursive specification over
\BTA\ in which no internal actions occur can be translated into a closed
\PGA\ term of which the behaviour is the solution of the finite guarded
recursive specification concerned.

Closed \PGA\ terms are considered to denote programs and therefore they
constitute an elementary program notation.
Closed \PGA\ terms are also called \PGA\ programs.

\section{A Hierarchy of Program Notations Rooted in Program Algebra}
\label{sect-hierarchy}

In~\cite{BL02a}, a hierarchy of program notations rooted in \PGA\ is
presented.
The program notations that appear in this hierarchy are \PGA, \PGLA,
\PGLB, \PGLC, \PGLD, \PGLDg, \PGLE, and \PGLS.
The most interesting ones are \PGLC, \PGLD\ and \PGLS.
\PGLC\ and \PGLD\ are close to existing assembly languages.
The main difference between them is that \PGLC\ has relative jump
instructions and \PGLD\ has absolute jump instructions.
\PGLS\ supports structured programming by offering a rendering of
conditional and loop constructs instead of (unstructured) jump
instructions.

\begin{figure}
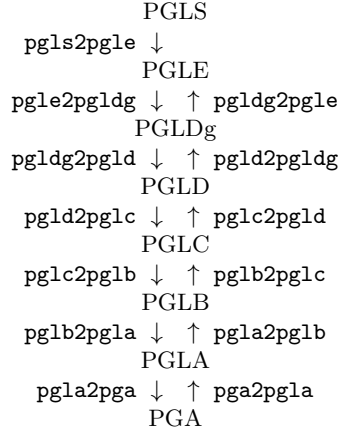

\centering
$
\begin{array}{@{}rcl@{}}
\multicolumn{3}{c}{\PGLS}                           \\
\pglspgle  & {} \downarrow \quad \phantom{\uparrow} \\
\multicolumn{3}{c}{\PGLE}                           \\
\pglepgldg & \downarrow \quad \uparrow & \pgldgpgle \\
\multicolumn{3}{c}{\PGLDg}                          \\
\pgldgpgld & \downarrow \quad \uparrow & \pgldpgldg \\
\multicolumn{3}{c}{\PGLD}                           \\
\pgldpglc  & \downarrow \quad \uparrow & \pglcpgld  \\
\multicolumn{3}{c}{\PGLC}                           \\
\pglcpglb  & \downarrow \quad \uparrow & \pglbpglc  \\
\multicolumn{3}{c}{\PGLB}                           \\
\pglbpgla  & \downarrow \quad \uparrow & \pglapglb  \\
\multicolumn{3}{c}{\PGLA}                           \\
\pglapga   & \downarrow \quad \uparrow & \pgapgla   \\
\multicolumn{3}{c}{\PGA}
\end{array}
$
\caption{Hierarchy of program notations rooted in \PGA}
\label{fig-hierarchy}
\end{figure}

For each of the program notations that appear in the hierarchy, except
\PGA, a function is given in~\cite{BL02a} by means of which each program
from that program notation is translated into a program from the first
program notation lower in the hierarchy that produces the same behaviour
on execution.
These functions are called projections.
Moreover, for each of the program notations that appear in the
hierarchy, except \PGLE\ and \PGLS, a function is given in~\cite{BL02a}
by means of which each program from that program notation is translated
into a program from the first program notation higher in the hierarchy
that produces the same behaviour on execution.
These functions are called embeddings.
There does not exist a function by which each \PGLE\ program is
translated into a \PGLS\ program that produces the same behaviour on
execution because \PGLS\ is strictly weaker than \PGLE.
The names of the projections and embeddings referred to above are given
in Figure~\ref{fig-hierarchy}.%

The program notations, projections, and embeddings referred to above are
defined in~\cite{BL02a}.
We refrain from giving all the definitions in question in this paper as
well, because most details of the program notations, projections, and
embeddings do not matter here.
The important point is that there exist a collection of well-defined
program notations rooted in an elementary program notation with
projections and embeddings between them.
Only \PGLD\ is actually used later on in this paper.
Therefore, \PGLD\ is reviewed below in Section~\ref{sect-PGLD}.

In Section~\ref{sect-polyadic-inseqs}, we will take special versions of
the above-mentioned program notations for the ones that may be used for
fragments.
Moreover, we will make use of the projections $\pglapga$ and in addition
the projections $\pglbpga$, $\pglcpga$, \ldots\ defined in the obvious
way by composition of projections given in~\cite{BL02a}.
In Section~\ref{sect-pg-synthesis}, we will make use of the embedding
$\pglcpgld$ and in addition the embedding $\pgapglc$ defined in the
obvious way by composition of embeddings given in~\cite{BL02a}.

\section{The Program Notation \PGLD}
\label{sect-PGLD}

In this section, we review the program notation \PGLD.
This program notation is reviewed because it will be used later on in
Sections~\ref{sect-example} and~\ref{sect-pg-synthesis}.

In \PGLD, like in \PGA, it is assumed that there is a fixed but
arbitrary finite set $\BInstr$ of \emph{basic instructions}.
Again, the intuition is that the execution of a basic instruction $a$
produces either $\True$ or $\False$ at its completion.

\PGLD\ has the following primitive instructions:
\begin{itemize}
\item
for each $a \in \BInstr$, a \emph{plain basic instruction} $a$;
\item
for each $a \in \BInstr$, a \emph{positive test instruction} $\ptst{a}$;
\item
for each $a \in \BInstr$, a \emph{negative test instruction} $\ntst{a}$;
\item
for each $l \in \Nat$, an \emph{absolute jump instruction}
$\ajmp{l}$.
\end{itemize}
\PGLD\ programs have the form $u_1;\ldots;u_k$, where $u_1,\ldots,u_k$
are primitive instructions of \PGLD.

The plain basic instructions, the positive test instructions, and the
negative test instructions are as in \PGA.
The effect of an absolute jump instruction $\ajmp{l}$ is that execution
proceeds with the $l$-th instruction of the program concerned.
If $\ajmp{l}$ is itself the $l$-th instruction, then $\ajmp{l}$ results
in inaction.
If $l$ equals $0$ or $l$ is greater than the length of the program, then
termination occurs.

The function $\pgldpga$ from the set of all \PGLD\ programs to the set
of all \PGA\ programs, which translates each \PGLD\ program into a \PGA\
program that produces the same behaviour on execution, can be defined
directly as follows:
\begin{ldispl}
\pgldpga(u_1 \conc \ldots \conc u_k) =
(\psi_1(u_1) \conc \ldots \conc \psi_k(u_k) \conc
 \halt \conc \halt)\rep\;,
\end{ldispl}%
where the auxiliary functions $\psi_j$ from the set of all primitive
instructions of \PGLD\ to the set of all primitive instructions of \PGA\
are defined as follows ($1 \leq j \leq k$):
\begin{ldispl}
\begin{aceqns}
\psi_j(\ajmp{l}) & = & \fjmp{l-j}       & \mif j \leq l \leq k\;, \\
\psi_j(\ajmp{l}) & = & \fjmp{k+2-(j-l)} & \mif 0   <  l   <  j\;, \\
\psi_j(\ajmp{l}) & = & \halt            & \mif l = 0 \lor l > k\;, \\
\psi_j(u)        & = & u
                    & \mif u\; \mathrm{is\;not\;a\;jump\;instruction}\;.
\end{aceqns}
\end{ldispl}%
The idea is that each backward jump can be replaced by a forward jump if
the entire program is repeated.
To enforce termination of the program after execution of its last
instruction if the last instruction is a plain basic instruction, a
positive test instruction or a negative test instruction,
$\halt \conc \halt$ is appended to
$\psi_1(u_1) \conc \ldots \conc \psi_k(u_k)$.

\section{Polyadic Instruction Sequences}
\label{sect-polyadic-inseqs}

In this section, we formalize a simple mechanism by which several
instruction sequence fragments can produce a joint behaviour.
The instruction sequence fragments are viewed as instruction sequences
that contain special instructions for switching over execution from one
fragment to another.
The instruction sequences in question are called polyadic instruction
sequences.

It is assumed that a special version of \PGLA, \PGLB, \PGLC, \PGLD,
\PGLDg, \PGLE\ or \PGLS\ is used for each polyadic instruction sequence.
Moreover, it is assumed that a collection of polyadic instruction
sequences between which execution can be switched takes the form of a
sequence, called a polyadic instruction sequence vector, in which each
polyadic instruction sequence is coupled with the program notation used
for it.

Our general view on the way of achieving a joint behaviour of the
polyadic instruction sequences in a polyadic instruction sequence vector
is as follows:
\begin{itemize}
\item
there can only be a single polyadic instruction sequence being executed
at any stage;
\item
the polyadic instruction sequence in question may make any polyadic
instruction sequence in the vector the one being executed;
\item
making another polyadic instruction sequence the one being executed is
effected by executing a special instruction for switching over
execution;
\item
any polyadic instruction sequence  can be taken for the one being
executed initially.
\end{itemize}

In addition to special instructions for switching over execution,
polyadic instruction sequences may contain two other kinds of special
instructions:
\begin{itemize}
\item
special instructions for putting instructions into instruction registers;
\item
special instructions of which the occurrences in a polyadic instruction
sequence are replaced by instructions contained in instruction registers
on making the polyadic instruction sequence the one being executed.
\end{itemize}
The special instructions of the latter kind serve as instruction
place-holders.
Their presence turns a polyadic instruction sequence into a
parameterized instruction sequence of which the parameters are filled in
each time it is made the one being executed.
This feature accounts for the use of the prefix polyadic.
Its merit is primarily that it allows for execution to proceed in effect
from different positions each time a polyadic instruction sequence is
loaded for execution.
An example of this is given in Section~\ref{sect-example}.

We take the line that different program notations can be used for
different polyadic instruction sequences in a polyadic instruction
sequence vector.
On making a polyadic instruction sequence in the vector the one being
executed, it is considered to be translated into a \PGAp\ program.

\PGAp\ is a variant of \PGA\ in which the above-mentioned special
instructions are incorporated.
In \PGAp, it is assumed that there is a fixed but arbitrary finite set
$\CBInstr$ of \emph{core basic instructions}.
In \PGAp, a basic instruction is either a core basic instruction or a
supplementary basic instruction.

\PGAp\ has the following \emph{core primitive instructions}:
\begin{itemize}
\item
for each $a \in \CBInstr$, a \emph{plain basic instruction} $a$;
\item
for each $a \in \CBInstr$, a \emph{positive test instruction} $\ptst{a}$;
\item
for each $a \in \CBInstr$, a \emph{negative test instruction} $\ntst{a}$;
\item
for each $l \in \Nat$, a \emph{forward jump instruction} $\fjmp{l}$;
\item
a \emph{termination instruction} $\halt$.
\end{itemize}
We write $\CPInstr$ for the set of all core primitive instructions.
The core primitive instructions of \PGAp\ are the counterparts of the
primitive instructions of \PGA.

\PGAp\ has the following \emph{supplementary basic instructions}:
\begin{itemize}
\item
for each $i \in \Nat$, a \emph{switch-over instruction} $\swo{i}$;
\item
for each $i \in \Nat$ and $u \in \CPInstr$, a \emph{put instruction}
$\puti{i}{u}$;
\item
for each $i \in \Nat$, a \emph{get instruction} $\geti{i}$.
\end{itemize}
We write $\SBInstr$ for the set of all supplementary basic instructions.
In the presence of a polyadic instruction sequence vector, a switch-over
instruction $\swo{i}$ is the instruction for switching over execution to
the $i$-th polyadic instruction sequence in the vector.
A put instruction $\puti{i}{u}$ is the instruction for putting
instruction $u$ in the instruction register with number $i$.
A get instruction $\geti{i}$ is the instruction of which each occurrence
in a polyadic instruction sequence is replaced by the contents of the
instruction register with number $i$ on switching over execution to that
polyadic instruction sequence.
If a get instruction is encountered in the polyadic instruction sequence
being executed, inaction occurs.

The supplementary basic instructions of \PGAp\ can be viewed as
built-in basic instructions.
However, as laid down below, supplementary basic instructions do not
occur in positive or negative test instructions.
Thus, the core primitive instructions and supplementary basic
instructions make up the primitive instructions of \PGAp.

\PGAp\ has the following constants and operators:
\begin{itemize}
\item
for each $u \in \CPInstr \union \SBInstr$, an \emph{instruction}
constant $u$\,;
\item
the binary \emph{concatenation} operator $\ph \conc \ph$\,;
\item
the unary \emph{repetition} operator $\ph\rep$\,.
\end{itemize}
The axioms of \PGAp\ are the same as the axioms of \PGA.

Suppose that in \PGA\ the restriction is dropped that $\BInstr$ must be
a finite set.
Then \PGAp\ can be viewed as the specialization of \PGA\ obtained by
taking the set $\CBInstr \union \SBInstr$ for $\BInstr$ and excluding
terms in which basic instructions from $\SBInstr$ occur in positive or
negative test instructions.
Henceforth, we actually drop the restriction that $\BInstr$ must be a
finite set.
This simplifies the definitions of the different program notations that
can be used for polyadic instruction sequences and also enables the use
of the functions $\pglapga$, $\pglbpga$, etcetera for translating
programs in those program notations into \PGAp\ programs.

The different program notations that can be used for polyadic
instruction sequences are \PGLAp, \PGLBp, \PGLCp, \PGLDp, \PGLDgp,
\PGLEp, and \PGLSp.
The set of all \PGLAp\ programs is the subset of the set of all \PGLA\
programs, taking the set $\CBInstr \union \SBInstr$ for $\BInstr$, in
which the basic instructions from $\SBInstr$ do not occur in positive or
negative test instructions.
The other program notations are defined similarly.
If the set $\CBInstr \union \SBInstr$ is taken for $\BInstr$, the
function $\pglapga$ translates each \PGLAp\ program into a \PGAp\
program that produces the same behaviour on execution.
Similar remarks apply to the other program notations.

A \emph{polyadic instruction sequence} is either a \PGLAp\ program, a
\PGLBp\ program, a \PGLCp\ program, a \PGLDp\ program, a \PGLDgp\
program, a \PGLEp\ program or a \PGLSp\ program.

A \emph{polyadic instruction sequence vector} is a sequence of pairs
consisting of a polyadic instruction sequence and a member of the set
$\set{A,B,C,D,Dg,E,S}$ of \emph{program notation indices}.

Let $\alpha$ be a polyadic instruction sequence vector,
let $P_1,\ldots,P_n$ and $c_1,\ldots,c_n$ be polyadic instruction
sequences and program notation indices, respectively, such that
$\alpha =
 \seq{\tup{P_1,c_1}} \concat \ldots \concat \seq{\tup{P_n,c_n}}$,%
\footnote
{We write $\seqof{D}$ for the set of all finite sequences with elements
 from set $D$.
 We use the following notation for finite sequences:
 $\emptyseq$ for the empty sequence,
 $\seq{d}$ for the sequence having $d$ as sole element,
 $\sigma \concat \sigma'$ for the concatenation of finite sequences
 $\sigma$ and $\sigma'$, and
 $\len(\sigma)$ for the length of finite sequence $\sigma$.}
and let $i \in [1,n]$.
Then we write $\pg(\alpha,i)$ and $\pgn(\alpha,i)$ for $P_i$ and $c_i$,
respectively.

Let $\alpha$ be a polyadic instruction sequence vector of length $n$,
and let $i \in [1,n]$.
Then program notation index $\pgn(\alpha,i)$ indicates which program
notation is used for polyadic instruction sequence $\pg(\alpha,i)$.
$A$ stands for \PGLAp, $B$ stands for \PGLBp, etcetera.
The program notation used is made explicit because it cannot always be
determined uniquely from the polyadic instruction sequence concerned,
whereas the behaviour that this polyadic instruction sequence produces
on execution may be different for each of the program notations in
question.

Below, we use special notation relating to the program notation indices.
Let $c$ be a program notation index.
Then we write $\prj_c$ for the projection $\pglapga$ if $c = A$,
the projection $\pglbpga$ if $c = B$, etcetera.

The set of instruction registers that contain an instruction and the
contents of each of those registers matter when a polyadic instruction
sequence is made the one being executed.
That makes us introduce the notion of an instruction register file state
and special notation relating to this notion.

An \emph{instruction register file state} is a function
$\funct{\sigma}{I}{\CPInstr}$, where $I$ is a finite subset of $\Nat$.

Let $p$ be a \PGAp\ program and $\sigma$ be an instruction register file
state.
Then we write $p[\sigma]$ for $p$ with, for all $i \in \dom(\sigma)$,
all occurrences of $\geti{i}$ in $p$ replaced by $\sigma(i)$.

Let $i,n \in \Nat$ be such that $1 \leq i \leq n$, let $\alpha$ be a
polyadic instruction sequence vector of length $n$, and let $\sigma$ be
an instruction register file state.
Then we write $\valid(\alpha,i,\sigma)$ to indicate that instructions of
the form $\geti{i}$ do not occur in
$\prj_{\pgn(\alpha,i)}(\pg(\alpha,i))[\sigma]$.

An obvious choice of the thread extraction operation of \PGAp\ is the
thread extraction operation of \PGA, taking the set
$\CBInstr \union \SBInstr$ for $\BInstr$, restricted to the set of
closed terms of \PGAp.
This thread extraction operation is considered not to be the proper one,
because it treats the supplementary basic instructions as arbitrary
basic instructions and thus disregards the fixed effects that they
produce on execution.
Moreover, this thread extraction operation requires that in \BTA\ the
restriction is dropped that $\BAct$ must be a finite set.

As regards the proper thread extraction for \PGAp, the idea is that it
yields, for each \PGAp\ program $P$, a function that determines, for
each polyadic instruction sequence vector $\alpha$, a finite guarded
recursive specification over \BTA\ that defines the thread that is the
joint behaviour of $P$ and the polyadic instruction sequences in
$\alpha$ in the case where $P$ is the polyadic instruction sequence
being executed initially.
Because this behaviour depends upon the set of instruction registers
that contain an instruction and the contents of each of those registers,
we need a thread extraction operation for each instruction register file
state.

In the thread extraction for \PGAp, it is assumed that $\gnl \in \IAct$.
The internal action $\gnl$ (generate and load) represents the internal
activity involved in switching over execution.
The internal actions $\Tau$ and $\gnl$ are distinguished because $\Tau$
is considered to represent a negligible internal activity, whereas
$\gnl$ is considered to represent a substantial internal activity.

For each instruction register file state $\sigma$, we introduce the
\emph{thread extraction} operation $\extrp{\ph}{\sigma}$.
These thread extraction operations are defined by the equations given in
Table~\ref{axioms-thread-extr-pp} (for $a \in \BInstr$, $l,i \in \Nat$,
$u \in \CPInstr \union \SBInstr$ and $v \in \CPInstr$)%
\begin{table}[!t]
\caption{Defining equations for thread extraction operations of \PGAp}
\label{axioms-thread-extr-pp}
\begin{eqntbl}
\begin{seqncol}
\extrp{a}{\sigma}(\alpha) = a \bapf \DeadEnd
\\
\extrp{a \conc x}{\sigma}(\alpha) = a \bapf \extrp{x}{\sigma}(\alpha)
\\
\extrp{\ptst{a}}{\sigma}(\alpha) = a \bapf \DeadEnd
\\
\extrp{\ptst{a} \conc x}{\sigma}(\alpha) =
\pcc{\extrp{x}{\sigma}(\alpha)}{a}
    {\extrp{\fjmp{2} \conc x}{\sigma}(\alpha)}
\\
\extrp{\ntst{a}}{\sigma}(\alpha) = a \bapf \DeadEnd
\\
\extrp{\ntst{a} \conc x}{\sigma}(\alpha) =
\pcc{\extrp{\fjmp{2} \conc x}{\sigma}(\alpha)}{a}
    {\extrp{x}{\sigma}(\alpha)}
\eqnsep
\extrp{\fjmp{l}}{\sigma}(\alpha) = \DeadEnd
\\
\extrp{\fjmp{0} \conc x}{\sigma}(\alpha) = \DeadEnd
\\
\extrp{\fjmp{1} \conc x}{\sigma}(\alpha) = \extrp{x}{\sigma}(\alpha)
\\
\extrp{\fjmp{l+2} \conc u}{\sigma}(\alpha) = \DeadEnd
\\
\extrp{\fjmp{l+2} \conc u \conc x}{\sigma}(\alpha) =
\extrp{\fjmp{l+1} \conc x}{\sigma}(\alpha)
\\
\extrp{\halt}{\sigma}(\alpha) = \Stop
\\
\extrp{\halt \conc x}{\sigma}(\alpha) = \Stop
\eqnsep
\extrp{\swo{i}}{\sigma}(\alpha) =
\gnl \bapf
\extrp{\prj_{\pgn(\alpha,i)}(\pg(\alpha,i))[\sigma]}{\sigma}(\alpha)
 & \mif 1 \leq i \leq n \land \valid(\alpha,i,\sigma)
\\
\extrp{\swo{i}}{\sigma}(\alpha) = \DeadEnd
 & \mif 1 \leq i \leq n \land \lnot \valid(\alpha,i,\sigma)
\\
\extrp{\swo{i}}{\sigma}(\alpha) = \Stop
 & \mif i = 0 \lor i > n
\\
\extrp{\swo{i} \conc x}{\sigma}(\alpha) =
\gnl \bapf
\extrp{\prj_{\pgn(\alpha,i)}(\pg(\alpha,i))[\sigma]}{\sigma}(\alpha)
 & \mif 1 \leq i \leq n \land \valid(\alpha,i,\sigma)
\\
\extrp{\swo{i} \conc x}{\sigma}(\alpha) = \DeadEnd
 & \mif 1 \leq i \leq n \land \lnot \valid(\alpha,i,\sigma)
\\
\extrp{\swo{i} \conc x}{\sigma}(\alpha) = \Stop
 & \mif i = 0 \lor i > n
\\
\extrp{\puti{i}{v}}{\sigma}(\alpha) = \Tau \bapf \DeadEnd
\\
\extrp{\puti{i}{v} \conc x}{\sigma}(\alpha) =
\Tau \bapf \extrp{x}{\sigma \owr \maplet{i}{v}}(\alpha)
\\
\extrp{\geti{i}}{\sigma}(\alpha) = \DeadEnd
\\
\extrp{\geti{i} \conc x}{\sigma}(\alpha) = \DeadEnd
\end{seqncol}
\end{eqntbl}
\end{table}
and the rule that $\extrp{\fjmp{l} \conc X}{\sigma}(\alpha) = \DeadEnd$
if $\fjmp{l}$ is the beginning of an infinite chain of forward jumps.

We can couple nominal indices as labels with some of the polyadic
instruction sequences in a polyadic instruction sequence vector.
This would permit the use of alternative switch-over instructions with
nominal indices instead of ordinal indices, like with the goto
instructions from \PGLDg.
In the notational style of~\cite{BB06a}, the form of those alternative
switch-over instructions would be $\swo{[i]}$.

\section{Example}
\label{sect-example}

To illustrate the mechanism formalized in
Section~\ref{sect-polyadic-inseqs}, we consider in this section the
splitting of a \PGLD\ program $P$ of $10000$ instructions into two
fragments.

We write $\nu_1(l)$ for the number of absolute jump instructions
$\ajmp{l'}$ with $l' > 5000$ from position $1$ up to position $l$ and
$\nu_2(l)$ for the number of absolute jump instructions $\ajmp{l'}$ with
$l' \leq 5000$ from position $5001$ up to position $l$.

The polyadic instruction sequence $P'$ corresponding to the first half
of $P$ is obtained from the first half of $P$ as follows:
\begin{itemize}
\item
the instruction $\geti{1}$ is prefixed to it;
\item
each absolute jump instruction $\ajmp{l}$ with $l \leq 5000$ is
replaced by the absolute jump instructions $\ajmp{l'}$, where
$l' = l + \nu_1(l) + 1$;
\item
each absolute jump instruction $\ajmp{l}$ with $l > 5000$ is replaced
by the instruction sequence $\puti{2}{\fjmp{l'}} \conc \swo{2}$, where
$l' = (l - 5000) + \nu_2(l - 5000)$;
\end{itemize}
and the polyadic instruction sequence $P''$ corresponding to the second
half of $P$ is obtained from the second half of $P$ as follows:
\begin{itemize}
\item
the instruction $\geti{2}$ is prefixed to it;
\item
each absolute jump instruction $\ajmp{l}$ with $l > 5000$ is
replaced by the absolute jump instructions $\ajmp{l'}$, where
$l' = (l - 5000) + \nu_2(l - 5000) + 1$;
\item
each absolute jump instruction $\ajmp{l}$ with $l \leq 5000$ is
replaced by the instruction sequence
$\puti{1}{\fjmp{l'}} \conc \swo{1}$, where $l' = l + \nu_1(l)$.
\end{itemize}
Notice that the positions occurring in jump instructions are adapted to
the prefixing of a get instruction to each half of $P$ and the
replacement of each jump instructions that gives rise to a jump into the
other half of $P$ by two instructions.

For any instruction register file state $\sigma$, we have that
$\extr{P}$ coincides with
$\extrp{\puti{1}{\fjmp{1}} \conc \swo{1}}{\sigma}
 (\seq{\tup{P',D}} \concat \seq{\tup{P'',D}})$
after the presence of the internal actions $\Tau$ and $\gnl$ in the
latter behaviour has been concealed.
In Section~\ref{sect-TSU}, we will introduce operators to conceal the
presence of internal actions.

In this section, we have illustrated by means of an example that
splitting a program into fragments is relatively simple.
In Section~\ref{sect-pg-synthesis}, we will show that synthesizing a
program from a collection of fragments is fairly complicated.

\section{Threads-Services Interaction and Abstraction}
\label{sect-TSU}

A thread may make use of services.
That is, it may perform certain actions for the purpose of having itself
affected by a service that takes them as commands to be processed.
The processing of an action may involve a change of state of the service
and at completion of the processing of the action the service returns a
reply value to the thread.
The reply value determines how the thread proceeds.
In this section, we review the use operators, which are concerned with
threads making such use of services.%
\footnote
{This version of the use mechanism was first introduced in~\cite{BM04c}.
 In later papers, it is also called thread-service composition.}
We also introduce abstraction operators.
They serve for concealment of the presence of internal actions, which
arise among other things from the use operators.
Both use operators and abstraction operators will be used later on in
Section~\ref{sect-pg-synthesis}.

It is assumed that a fixed but arbitrary finite set $\Foci$ of
\emph{foci} and a fixed but arbitrary finite set $\Meth$ of
\emph{methods} have been given.
Each focus plays the role of a name of a service provided by the
execution environment that can be requested to process a command.
Each method plays the role of a command proper.
For the set $\BAct$ of basic actions, we take the set
$\set{f.m \where f \in \Foci, m \in \Meth}$.
Performing a basic action $f.m$ is taken as making a request to the
service named $f$ to process command $m$.

A \emph{service} $H$ consists of
\begin{itemize}
\item
a set $S$ of \emph{states};
\item
an \emph{effect} function $\funct{\eff}{\Meth \cp S}{S}$;
\item
a \emph{yield} function
$\funct{\yld}{\Meth \cp S}{\set{\True,\False,\Blocked}}$;
\item
an \emph{initial state} $s_0 \in S$;
\end{itemize}
satisfying the following condition:
\begin{ldispl}
\Forall{m \in \Meth, s \in S}
{(\yld(m,s) = \Blocked \Implies
  \Forall{m' \in \Meth}{\yld(m',\eff(m,s)) = \Blocked})}\;.
\end{ldispl}%
The set $S$ contains the states in which the service may be, and the
functions $\eff$ and $\yld$ give, for each method $m$ and state $s$, the
state and reply, respectively, that result from processing $m$ in state
$s$.
By the condition imposed on services, once the service has returned
$\Blocked$ as reply, it keeps returning $\Blocked$ as reply.

Let $H = \tup{S,\eff,\yld,s_0}$ be  a service and let $m \in \Meth$.
Then
the \emph{derived service} of $H$ after processing $m$, written
$\derive{m}H$, is the service $\tup{S,\eff,\yld,\eff(m,s_0)}$; and
the \emph{reply} of $H$ after processing $m$, written $H(m)$, is
$\yld(m,s_0)$.

When a thread makes a request to the service to process $m$:
\begin{itemize}
\item
if $H(m) \neq \Blocked$, then the request is accepted, the reply is
$H(m)$, and the service proceeds as $\derive{m}H$;
\item
if $H(m) = \Blocked$, then the request is rejected and the service
proceeds as a service that rejects any request.
\end{itemize}

We introduce the sort $\Serv$ of \emph{services} and,
for each $f \in \Foci$, the binary \emph{use} operator
$\funct{\use{\ph}{f}{\ph}}{\Thr \cp \Serv}{\Thr}$.
The axioms for these operators are given in Table~\ref{axioms-tsu}.%
\begin{table}[!t]
\caption{Axioms for use operators}
\label{axioms-tsu}
\begin{eqntbl}
\begin{saxcol}
\use{\Stop}{f}{H} = \Stop                            & & \axiom{TSU1} \\
\use{\DeadEnd}{f}{H} = \DeadEnd                      & & \axiom{TSU2} \\
\use{(\iota \bapf x)}{f}{H} =
                         \iota \bapf (\use{x}{f}{H}) & & \axiom{TSU3} \\
\use{(\pcc{x}{g.m}{y})}{f}{H} =
\pcc{(\use{x}{f}{H})}{g.m}{(\use{y}{f}{H})}
 & \mif f \neq g                                       & \axiom{TSU4} \\
\use{(\pcc{x}{f.m}{y})}{f}{H} =
\Tau \bapf (\use{x}{f}{\derive{m}H})
 & \mif H(m) = \True                                   & \axiom{TSU5} \\
\use{(\pcc{x}{f.m}{y})}{f}{H} =
\Tau \bapf (\use{y}{f}{\derive{m}H})
 & \mif H(m) = \False                                  & \axiom{TSU6} \\
\use{(\pcc{x}{f.m}{y})}{f}{H} = \DeadEnd
 & \mif H(m) = \Blocked                                & \axiom{TSU7}
\end{saxcol}
\end{eqntbl}
\end{table}
In this table, $\iota$ stands for an arbitrary member of $\IAct$, $f$
and $g$ stand for arbitrary foci from $\Foci$, and $m$ stands for an
arbitrary method from~$\Meth$.

Intuitively, $\use{p}{f}{H}$ is the thread that results from processing
all basic actions performed by thread $p$ that are of the form $f.m$ by
service $H$.
When a basic action of the form $f.m$ performed by thread $p$ is
processed by service $H$, it is turned into the internal action $\Tau$
and postconditional composition is removed in favour of action prefixing
on the basis of the reply value produced.
This intuition is covered by axioms TSU5 and TSU6.
Axioms TSU3 and TSU4 express that internal actions and basic actions of
the form $g.m$ with $f \neq g$ are not processed.
Axiom TSU7 expresses that inaction occurs when a basic action to be
processed is not accepted.

Let $T$ stand for either \BTA, \BTA+\REC\ or \BTA+\REC+\AIP.
Then we will write $T$+\TSU\ for $T$, taking the set
$\set{f.m \where f \in \Foci, m \in \Meth}$ for $\BAct$, extended with
the use operators and the axioms from Table~\ref{axioms-tsu}.

For each $\iota \in \IAct$, we introduce the unary \emph{abstraction}
operator $\funct{\abstr{\iota}}{\Thr}{\Thr}$ to conceal the presence of
internal action $\iota$ in the case where its presence does not matter.
The axioms for the abstraction operators are given in
Table~\ref{axioms-abstr}.%
\begin{table}[!t]
\caption{Axioms for abstraction operators}
\label{axioms-abstr}
\begin{eqntbl}
\begin{saxcol}
\abstr{\iota}(\Stop) = \Stop                          & & \axiom{TT1} \\
\abstr{\iota}(\DeadEnd) = \DeadEnd                    & & \axiom{TT2} \\
\abstr{\iota}(\iota \bapf x) = \abstr{\iota}(x)       & & \axiom{TT3} \\
\abstr{\iota}(\pcc{x}{a}{y}) =
\pcc{\abstr{\iota}(x)}{a}{\abstr{\iota}(y)}
 & \mif a \neq \iota                                    & \axiom{TT4} \\
\AND{n \geq 0} \abstr{\iota}(\proj{n}{x}) = \abstr{\iota}(\proj{n}{y})
 \Implies \abstr{\iota}(x) = \abstr{\iota}(y)         & & \axiom{TT5}
\end{saxcol}
\end{eqntbl}
\end{table}
In this table, $a$ stands for an arbitrary action from $\Act$.

A main difference between the version of the use mechanism introduced
here and the version of the use mechanism introduced in~\cite{BP02a} is
that the former version does not incorporate abstraction and the latter
version incorporates abstraction.

Let $T$ stand for either \BTA, \BTA+\REC, \BTA+\REC+\AIP, \BTA+\TSU,
BTA+\REC+\TSU\ or \BTA+\REC+\AIP+\TSU.
Then we will write $T$+\ABSTR\ for $T$ extended with the abstraction
operators and the axioms from Table~\ref{axioms-abstr}.

For each $\iota \in \IAct$, the equation
$\abstr{\iota}(\iota^\omega) = \DeadEnd$
is derivable from the axioms of \BTA+\REC+\AIP+\ABSTR.

\section{Instruction Register File Services}
\label{sect-IRF}

In this section, we describe services that make up register files with a
finite set of registers that can contain instructions from a finite set
of core primitive instructions.
These services will be used in Section~\ref{sect-pg-synthesis}.

It is assumed that a fixed but arbitrary set $\Reg \subseteq \Nat$ such
that $\Reg = [1,n]$ for some $n \in \Nat$ and a fixed but arbitrary
finite set $\Instr \subseteq \CPInstr$ have been given.
The set $\Reg$ is considered to contain the positions of the registers
in the instruction register file and the set $\Instr$ is considered to
contain the instructions that can be put in those registers.

We write $\nm{IRFS}$ for the set
$\Union{\Reg' \subseteq \Reg} \mapof{\Reg'}{\Instr}$.
The members of $\nm{IRFS}$ are considered to be the possible instruction
register file states.
It is assumed that a fixed but arbitrary bijection
$\funct{\theta}{\nm{IRFS}}{[1,\card(\nm{IRFS})]}$ has been given.

The instruction register file services accept the following methods:
\begin{itemize}
\item
for each $i \in \Reg$ and $u \in \Instr$,
a \emph{register put method} $\putr{:}i{:}u$;
\item
for each
$j \in \rng(\theta)$,
a \emph{register file test method} $\eqr{:}j$.
\end{itemize}
We write $\Meth_\irf$ for the set
$\set{\putr{:}i{:}u \where i \in \Reg \land u \in \Instr} \union
 \set{\eqr{:}j \where j \in \rng(\theta)}$.

It is assumed that $\Meth_\irf \subseteq \Meth$.

The methods accepted by instruction register file services can be
explained as follows:
\begin{itemize}
\item
$\putr{:}i{:}u$\,:
the contents of register $i$ becomes instruction $u$ and the reply is
$\True$;
\item
$\eqr{:}j$\,:
if the state of the instruction register file equals $\theta^{-1}(j)$,
then nothing changes and the reply is $\True$; otherwise nothing changes
and the reply is $\False$.
\end{itemize}

Let $s \in \nm{IRFS} \union \set{\undef}$,
where $\undef \notin \nm{IRFS}$.
Then the \emph{instruction register file service} with initial state
$s$, written $\IRF_s$, is the service $\tup{\nm{IRFS},\eff,\yld,s}$,
where the functions $\eff$ and $\yld$ are defined as follows
($\sigma \in \nm{IRFS} \union \set{\undef}$):%
\footnote
{We use the following notation for functions:
 $\emptymap$ for the empty function;
 $\maplet{d}{r}$ for the function $f$ with $\dom(f) = \set{d}$ such that
 $f(d) = r$; and
 $f \owr g$ for the function $h$ with $\dom(h) = \dom(f) \union \dom(g)$
 such that, for all $d \in \dom(h)$,
 $h(d) = f(d)$ if $d \not\in \dom(g)$ and $h(d) = g(d)$ otherwise.}%
\begin{ldispl}
\begin{geqns}
\eff(\putr{:}i{:}u,\sigma) = \sigma \owr \maplet{i}{u}\;,
\\
\eff(\eqr{:}j,\sigma)  = \sigma\;,
\\ {} \\
\eff(m,\sigma)      = \undef  \qquad \mif m \not\in \Meth_\irf\;,
\\
\eff(m,\undef) = \undef\;,
\end{geqns}
\qquad
\begin{gceqns}
\yld(\putr{:}i{:}n,\sigma) = \True\;,
\\
\yld(\eqr{:}j,\sigma) = \True  & \mif \theta(\sigma) = j\;,
\\
\yld(\eqr{:}j,\sigma) = \False & \mif \theta(\sigma) \neq j\;,
\\
\yld(m,\sigma)      = \Blocked     & \mif m \not\in \Meth_\irf\;,
\\
\yld(m,\undef) = \Blocked\;.
\end{gceqns}
\end{ldispl}%

\section{Program Synthesis}
\label{sect-pg-synthesis}

In order to establish a connection between collections of instruction
sequence fragments and instruction sequences, we show in this section
that, for each possible joint behaviour of a collection of instruction
sequence fragments, a single instruction sequence can be synthesized
from the collection that produces on execution essentially the behaviour
in question by making use of an instruction register file service.
More precisely, we show that, for each \PGAp\ program $P$ and polyadic
instruction sequence vector $\alpha$, a \PGA\ program $P'$ can be
synthesized from $P$ and $\alpha$ such that, for all relevant
instruction register file states $\sigma$,
$\use{\extr{P'}}{\irf}{\IRF_\sigma}$ coincides with
$\extrp{P}{\sigma}(\alpha)$ after the presence of the internal actions
$\Tau$ and $\gnl$ has been concealed.

Let $P$ be a \PGAp\ program and $\alpha$ be a polyadic instruction
sequence vector.
The general idea is that:
\begin{itemize}
\item
each polyadic instruction sequence in $\alpha$ is translated into a
\PGAp\ program and an appropriate finite collection of instances of this
\PGAp\ program in which occurrences of get instructions are replaced by
core primitive instructions is generated;
\item
$P$ and all the generated programs are translated into \PGLCp\ programs
and these \PGLCp\ programs are concatenated;
\item
the resulting \PGLCp\ program is translated into a \PGLDp\ program and
this program is translated into a \PGLD\ program by replacing all
occurrences of the supplementary instructions by core primitive
instructions as follows:
\begin{itemize}
\item
a switch-over instruction $\swo{i}$ is replaced by an absolute jump
instruction whose effect is a jump to the beginning of an appended
instruction sequence whose execution leads, after the state of the
instruction register file has been found by a linear search, to a jump
to the beginning of the right instance of the \PGAp\ program that
corresponds to the $i$th polyadic instruction sequence in $\alpha$;
\item
a put instruction $\puti{i}{u}$ is simply replaced by the plain basic
instruction $\irf.\putr{:}i{:}u$;
\item
a get instruction $\geti{i}$ is simply replaced by the absolute jump
instruction whose effect is a jump to the position of the instruction
itself.
\end{itemize}
\end{itemize}
A collection of instances of the \PGAp\ program corresponding to a
polyadic instruction sequence in $\alpha$ is considered appropriate if
it includes all instances that may become the one being executed.
$P$ and all the generated programs are translated into \PGLCp\ programs
because \PGLCp\ programs are relocatable: they can be concatenated
without disturbing the meaning of jump instructions.
The \PGLCp\ program resulting from the concatenation is translated into
a \PGLDp\ program before the supplementary instructions are replaced
because the replacement of a switch-over instruction by an absolute jump
instruction is simpler than its replacement by a relative jump
instruction.

Following the general idea outlined above, we define a function
$\pgappgld$ that yields, for each \PGAp\ program $P$, a function that
gives, for each polyadic instruction sequence vector $\alpha$, a PGLD\
program $P'$ such that, for each relevant instruction register file
service state $\sigma$, $\use{\extr{\pgldpga(P')}}{\irf}{\IRF_\sigma}$
coincides with $\extrp{P}{\sigma}(\alpha)$ after the presence of the
internal actions $\Tau$ and $\gnl$ has been concealed.

The function $\pgappgld$ from the set of all \PGAp\ programs to the set
all functions from the set of all polyadic instruction sequence vectors
to the set of all PGLD\ programs is defined as follows:
\begin{ldispl}
\pgappgld(x)(\alpha) = \\ \quad
\translate(\pglcpgld(\expand(x)(\alpha))) \conc {} \\ \quad
\ptst{\irf.\eqr{:}1} \conc \ajmp{l_{1,1}} \conc \ldots \conc
\ptst{\irf.\eqr{:}n'} \conc \ajmp{l_{1,n'}} \conc {}
\\ \qquad \vdots  \\ \quad
\ptst{\irf.\eqr{:}1} \conc \ajmp{l_{n,1}} \conc \ldots \conc
\ptst{\irf.\eqr{:}n'} \conc \ajmp{l_{n,n'}}\;,
\end{ldispl}%
where $n = \len(\alpha)$, $n' = \max(\rng(\theta))$,
the function $\expand$ from the set of all \PGAp\ programs to the
set all functions from the set of all polyadic instruction sequence
vectors to the set of all PGLCp\ programs is defined as follows:
\begin{ldispl}
\expand(x)(\alpha) = \\ \quad
\pgapglc(x) \conc {} \\ \quad
\pgapglc(\generate(\alpha,1,\theta^{-1}(1))) \conc \ldots \conc
\pgapglc(\generate(\alpha,1,\theta^{-1}(n'))) \conc {}
\\ \qquad \vdots \\ \quad
\pgapglc(\generate(\alpha,n,\theta^{-1}(1))) \conc \ldots \conc
\pgapglc(\generate(\alpha,n,\theta^{-1}(n')))\;,
\end{ldispl}%
\hspace*{2.12em}
\begin{minipage}{32.3em}
where $n = \len(\alpha)$, $n' = \max(\rng(\theta))$, and the function
$\generate$ from the set of all polyadic instruction sequence vectors,
the set of all natural numbers and the set of all instruction register
file states to the set of all \PGAp\ programs is defined as follows:
\end{minipage}
\pagebreak[2]
\begin{ldispl}
\begin{aceqns}
\generate(\alpha,i,\sigma) & = &
\prj_{\pgn(\alpha,i)}(\pg(\alpha,i))[\sigma]
 & \mif 1 \leq i \leq \len(\alpha) \land
        \valid(\alpha,i,\sigma)\;, \\
\generate(\alpha,i,\sigma) & = & \fjmp{0}
 & \mif 1 \leq i \leq \len(\alpha) \land
        \lnot \valid(\alpha,i,\sigma)\;, \\
\generate(\alpha,i,\sigma) & = & \halt
 & \mif i = 0 \lor i > \len(\alpha)\;, \\
\end{aceqns}
\end{ldispl}%
the function $\translate$ from the set of all \PGLDp\ programs to the
set of all \PGLD\ programs is defined as follows:
\begin{ldispl}
\translate(u_1 \conc \ldots \conc u_k) =
\psi_1(u_1) \conc \ldots \conc \psi_1(u_k)\;,
\end{ldispl}%
\hspace*{2.12em}
\begin{minipage}{32.3em}
where the functions $\psi_j$ from the set of all primitive instructions
of \PGLDp\ to the set of all primitive instructions of \PGLD\ are
defined as follows ($1 \leq j \leq k$):
\end{minipage}
\begin{ldispl}
\begin{aceqns}
\psi_j(\swo{i})  & = & \ajmp{l_i}
 & \mif 1 \leq i \leq \len(\alpha)\;, \\
\psi_j(\swo{i})  & = & \halt
 & \mif i = 0 \lor i > \len(\alpha)\;, \\
\psi_j(\puti{i}{u})  & = & \irf.\putr{:}i{:}u\;, \\
\psi_j(\geti{i})  & = & \ajmp{j}\;, \\
\psi_j(u) & = & u
 & \mif u\; \mathrm{is\;not\;a\;supplementary\;basic\;instruction}\;,
\end{aceqns}
\end{ldispl}%
\hspace*{2.12em}
\begin{minipage}{32.3em}
where for each $i \in [1,\len(\alpha)]$:
\end{minipage}
\begin{ldispl}
\begin{aeqns}
l_i & = &
\len(\pgapglc(x))
\\ & & {} +
\displaystyle\sum_{h \in [1,\len(\alpha)],h' \in \rng(\theta)}
 \len(\pgapglc(\prj_{\pgn(\alpha,h)}(\pg(\alpha,h))[\theta^{-1}(h')]))
\\ & & {} +
2 \mul \max(\rng(\theta)) \mul (i-1)\;,
\end{aeqns}
\end{ldispl}%
and for each $i \in [1,\len(\alpha)]$ and $j \in \rng(\theta)$:
\begin{ldispl}
\begin{aeqns}
l_{i,j} & = &
\len(\pgapglc(x))
\\ & & {} +
\displaystyle\sum_{h \in [1,i-1],h' \in \rng(\theta)}
 \len(\pgapglc(\prj_{\pgn(\alpha,h)}(\pg(\alpha,h))[\theta^{-1}(h')]))
\\ & & {} + \hsp{1.72}
\displaystyle\sum_{h' \in [1,j-1]}
 \len(\pgapglc(\prj_{\pgn(\alpha,i)}(\pg(\alpha,i))[\theta^{-1}(h')]))\;.
\end{aeqns}
\end{ldispl}%

The following theorem states rigorously that, for any \PGAp\ program
$P$ and polyadic instruction sequence vector $\alpha$, for all relevant
instruction register file states $\sigma$,
$\use{\extr{\pgldpga(\pgappgld(P)(\alpha))}}{\irf}{\IRF_\sigma}$
coincides with $\extrp{P}{\sigma}(\alpha)$ after the presence of the
internal actions $\Tau$ and $\gnl$ has been concealed.
\begin{theorem}
\label{theorem-pg-synthesis}
Let $P$ be a \PGAp\ program and $\alpha$ be a polyadic instruction
sequence vector, and let $n$ be the highest number occurring in
instructions of the form $\puti{i}{u}$ or $\geti{i}$ in $P$ or $\alpha$.
Take the interval $[1,n]$ for $\Reg$ and the set of all core primitive
instructions occurring in instructions of the form $\puti{i}{u}$ in $P$
or $\alpha$ for $\Instr$, and let $\sigma \in \nm{IRFS}$.
Then
$\abstr{\Tau}
  (\use{\extr{\pgldpga(\pgappgld(P)(\alpha))}}{\irf}{\IRF_\sigma}) =
 \abstr{\Tau}(\abstr{\gnl}(\extrp{P}{\sigma}(\alpha)))$.
\end{theorem}
\begin{proof}
The proof of Theorem~\ref{theorem-pg-synthesis} follows the same
line as the proof of Theorem~1 from~\cite{BM07f} given in that paper.
Here, we give only a brief outline of the proof of the current theorem.

The proof of this theorem proceeds as follows:
(i)~we give a set $T$ of closed terms of sort $\Thr$ with
$\abstr{\Tau}(\abstr{\gnl}(\extrp{P}{\sigma}(\alpha))) \in T$,
a set $T'$ of closed terms of sort $\Thr$ with
$\abstr{\Tau}
  (\use{\extr{\pgldpga(\pgappgld(P)(\alpha))}}{\irf}{\IRF_\sigma}) \in
 T'$, and
a bijection $\funct{\beta}{T}{T'}$;
(ii)~we show that there exists a set $E$ consisting of one derivable
equation $p = p'$ for each $p \in T$ such that, for all equations
$p = p'$ in $E$:
\begin{itemize}
\item
$\beta(p) = p''$ is a derivable equation if $p''$ is $p'$ with, for all
$q \in T$, all occurrences of $q$ in $p'$ replaced by $\beta(q)$;
\item
$p' \in T$ only if $p'$ can be rewritten to a $q' \not\in T$ using the
equations in $E$ from left to right.
\end{itemize}
This means that
$\abstr{\Tau}(\abstr{\gnl}(\extrp{P}{\sigma}(\alpha)))$
and
$\abstr{\Tau}
  (\use{\extr{\pgldpga(\pgappgld(P)(\alpha))}}{\irf}{\IRF_\sigma})$
denote solutions of the same guarded recursive specification.
Because guarded recursive specifications have unique solutions, it
follows immediately that
$\abstr{\Tau}
  (\use{\extr{\pgldpga(\pgappgld(P)(\alpha))}}{\irf}{\IRF_\sigma}) =
 \abstr{\Tau}(\abstr{\gnl}(\extrp{P}{\sigma}(\alpha)))$.
\qed
\end{proof}
In the proof outlined above, an apposite indexing of the closed terms in
the sets $T$ and $T'$ facilitates the definition of the bijection
$\beta$.
Yet, this definition is much more complicated than the definition of the
bijection needed in the proof from~\cite{BM07f} referred to.

The definition of $\pgappgld$ shows that the synthesis of single
instruction sequences from collections of instruction sequence fragments
is fairly complicated.

\section{Conclusions}
\label{sect-concl}

We have given a formalization of a simple mechanism by which several
instruction sequence fragments can produce a joint behaviour.
Thread extraction provides the possible joint behaviours of a collection
of instruction sequence fragments.
We have shown that, for each possible joint behaviour of a collection of
instruction sequence fragments, a single instruction sequence can be
synthesized from the collection that produces on execution essentially
the behaviour in question by making use of a service.
This program synthesis is reminiscent of the service-based variant of
projection semantics for program notations used in~\cite{BM07e}.
The program synthesis shows that it is a non-trivial matter to explain
by means of a translation into a single instruction sequence what takes
place on execution of a collection of instruction sequence fragments.

In this paper, an instruction sequence fragment in a collection of
instruction sequence fragments has two attributes: an ordinal index
(position) and a program notation index.
We have also mentioned that a nominal index (label) could be an optional
attribute.
Many other attributes that are relevant in practice can be imagined,
e.g.\ modification date, author, tester, owner, user, and security
level.
In this paper, we have restricted ourselves to attributes that are
indispensable for a theoretical understanding.

The question arises whether all aspects of the mechanism formalized in
this paper can be dealt with at the level of threads.
This issue has been investigated in~\cite{BM08f}.
It happens that not all aspects can be dealt with at the level of
threads.
In particular, the ability to replace special instructions in an
instruction sequence fragment by different ordinary instructions every
time execution is switched over to that fragment cannot be dealt with at
the level of threads.
Threads turn out to be too abstract to deal with the mechanism in full.

It is sometimes suggested that program algebra is closely related to
Kleene algebra~\cite{Koz94a} and omega algebra~\cite{Coh00a} -- an
extension of Kleene algebra with an infinite iteration operator.
In program algebra, programs are considered at a more concrete level
than in Kleene algebra and omega algebra.
For instance, Kleene algebra and omega algebra are not concerned with
instruction sequences, and program algebra is not concerned with
non-determinism.

\bibliographystyle{spmpsci}
\bibliography{TA}

\end{document}